\documentclass[aps,pra,twocolumn,superscriptaddress,merge,showpacs]{revtex4-1}
\usepackage[sort&compress]{natbib}
\usepackage{amsmath} % Include to use align command
\usepackage{graphicx}% Include figure files
\begin{document}

\title{The Generation of Super-Resolving Single-Photon Path-Entangled State}

\author{Wei Feng}
\affiliation{Beijing Computational Science Research Center, Beijing 100084, China}
\author{Kebei Jiang}
\affiliation{Hearne Institute for Theoretical Physics and Department of Physics and Astronomy \\
Louisiana State University, Baton Rouge, LA 70803 USA}
\author{Michelle L.-J. Lollie}
\affiliation{Hearne Institute for Theoretical Physics and Department of Physics and Astronomy \\
Louisiana State University, Baton Rouge, LA 70803 USA}
\affiliation{Department of Physics and Optical Engineering \\ 
Rose-Hulman Institute of Technology, Terre Haute, IN 47803 USA}
\author{M. Suhail Zubairy}
\affiliation{Computational Science Research Center, Beijing 100084, China}
\affiliation{Institute for Quantum Science and Engineering and Department of Physics and Astronomy\\
Texas A$\&$M University, College Station, TX 77843 USA}
\author{Jonathan P. Dowling}
\email{jdowling@lsu.edu}
\affiliation{Computational Science Research Center, Beijing 100084, China}
\affiliation{Hearne Institute for Theoretical Physics and Department of Physics and Astronomy \\
Louisiana State University, Baton Rouge, LA 70803 USA}

\date{\today}

\begin{abstract}

In this paper, we propose two protocols for generating super-resolving \textit{single-photon} path-entangled states from general maximally path-entangled N00N states. We also show that both protocols generate the desired state with different probabilities depending on the type of detectors being used. Such super-resolving single-photon path-entangled states preserve high resolving power but lack the requirement of a multi-photon absorbing resist, which makes this state a perfect candidate for quantum lithography.

\end{abstract}
%\pacs{42.50.St, 42.50.Dv, 03.65.Ud, 42.50.Lc}

\maketitle

%----------------------------------------------------------------------------------------

\section*{Introduction}

With the promising ability to beat the Rayleigh diffraction limit, Quantum lithography has drawn a great amount of attention ever since it was first proposed by Boto \textit{et al.} in 2000. The original proposal and the experiment  realizing quantum lithography \cite{Boto2000, Kok2001, D'Angelo2001} exploits the path-entanglement of an ensemble of $N$ photons whose de Broglie wavelengths are effectively $N$ times smaller than that of a single photon. However, one of the difficulties of such a scheme is that the arriving quantum-correlated photons are not always concentrated at the same absorption spot \cite{Steuernagel2004}. Moreover, an $N$-photon absorption process requires a multi-photon absorbing resist \cite{Hemmer2006, Sun2007} which limits their utility. Therefore, instead of utilizing photon entanglement, several other approaches with non-quantum states of light \cite{Wang2004, Bentley2004, Peer2004, Kiffner2008, Liao2010, Gorshkov2008, Ge2013} have been developed, all of which require either a nonlinearity material or resonant field-atom interaction.     

Here we show that it is possible to implement quantum lithography via a refined approach with entangled optical fields, where a super-resolving \textit{single-photon} path-entangled state is generated by reducing the photon number of an $N$ photon path-entangled state but preserving its $N$-fold resolving power. The aforementioned difficulties of the need for a multi-photon absorbing can be automatically resolved with the application of this type of super-resolving single-photon path-entangled state. 

An $N$ photon maximally path-entangled state, which is also known as the N00N state, in a two-path interferometer with a $\phi$ phase shift in one path is defined as \cite{Dowling2008} 
\begin{align}
\vert N::0 \rangle^{N\phi} \equiv\frac{1}{2}\left( \vert N,0 \rangle + e^{i N \phi} \vert 0,N\rangle \right).
\label{Eq:N00N}
\end{align}
Recently, a ``High-N00N'' state with $N$ up to five has been generated in the lab \cite{Afek2010}. The super-resolving power of the N00N state comes from the $N$-fold relative phase between $\vert N,0 \rangle$ and $\vert 0,N \rangle$. Consequently, a super-resolving single-photon path-entangled state, which refers to a N00N state with single photon number but original super-resolving power, would have the form
\begin{align}
\vert 1::0 \rangle^{N\phi} \equiv\frac{1}{2}\left( \vert 1,0 \rangle + e^{i N \phi} \vert 0,1\rangle \right).
\label{Eq:1001}
\end{align} 
For the sake of later calculation, we also define a more general N00N state whose super-resolving power is different from its photon number as 
\begin{align}
\vert N::0 \rangle^{M\phi} \equiv\frac{1}{2}\left( \vert N,0 \rangle + e^{i M \phi} \vert 0,N\rangle \right).
\label{Eq:N00M}
\end{align}
The purpose of this paper is to show how to produce states of Eq.~(\ref{Eq:1001}) from states of Eq.~(\ref{Eq:N00N}) by generating sequence of states of Eq.~(\ref{Eq:N00M}), with only linear quantum optical elements.

\begin{figure}[h] %
\centering %
\includegraphics[width=0.8\linewidth]{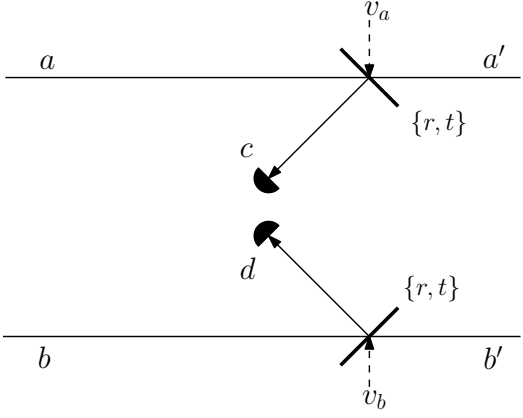} %
\caption{Two beamsplitters with the same reflectance $\rho=r^2$ are employed to reduce the number of the photons being transmitted. The existence of detector $c$ and $d$ reveals which-way information of the reflected photons and the incoming N00N state in mode $a$ and $b$ collapse into a separable state in mode $a'$ or $b'$. Therefore the relative phase is lost and the phase-resolving power is destroyed.} %
\label{Fig:whichway} %
\end{figure}

\section{Which-way information and Quantum Eraser}

In this first section we show that the acquisition of which-way information destroys the phase-resolving power of a maximally path-entangled state. For pedagogical reasons, we start with a $\vert 2::0 \rangle^{2\phi}$ as the input state before the beamsplitters in Fig.~\ref{Fig:whichway}. Assuming the two beamsplitters have same transmittance $t$ and reflectance $r$, and from the standard quantum beamsplitter transformation \cite{Gerry2005}
\begin{align}
\hat{a}^\dagger = t \hat{a'}^\dagger + i r \hat{c}^\dagger,~~~\hat{b}^\dagger = t \hat{b'}^\dagger + i r \hat{d}^\dagger,
\end{align}
it is straightforward to calculate the probabilities of transmitting all possible states to $a'$ and $b'$ and the result is shown in Table~\ref{Tab:whichway}. Notice that because of the availability of the which-way information of the incoming photon, the chance of transmitting $\vert 1::0 \rangle^{2\phi}$ is zero.

\begin{table}[h]
\begin{center}
\begin{tabular}{ | c | c | c |}
    \hline
          detector state       &       transmitted state        & probability \\ \hline
     $\vert 0,0 \rangle_{d,c}$ &   $\frac{t^2}{\sqrt{2}} \left( \vert 2,0 \rangle_{a',b'} + e^{i 2 \phi} \vert 0,2 \rangle_{a',b'} \right)$   & $  t^4 $  \\ \hline
     $\vert 0,1 \rangle_{d,c}$ &   $itr\vert 1,0 \rangle_{a',b'}$              & $ t^2 r^2 $ \\ \hline
     $\vert 0,2 \rangle_{d,c}$ &   $-\frac{r^2}{\sqrt{2}}\vert 0,0 \rangle_{a',b'}$    & $ \frac{1}{2} r^4 $  \\ \hline
     $\vert 1,0 \rangle_{d,c}$ &   $e^{i2\phi}itr\vert 0,1 \rangle_{a',b'}$                        & $ t^2 r^2 $ \\ \hline
     $\vert 2,0 \rangle_{d,c}$ &   $-e^{i2\phi}\frac{r^2}{\sqrt{2}}\vert 0,0 \rangle_{a',b'}$              & $ \frac{1}{2} r^4 $ \\ \hline
\end{tabular}
\caption{Probabilities of detecting all possible states at $c$ and $d$ for a $\vert 2::0 \rangle^{2\phi}$ input and accessible which-way information, as shown in Fig.~\ref{Fig:whichway}.}
\end{center}
\label{Tab:whichway}
\end{table}

\begin{figure}[th] %
\centering %
\includegraphics[width=0.8\linewidth]{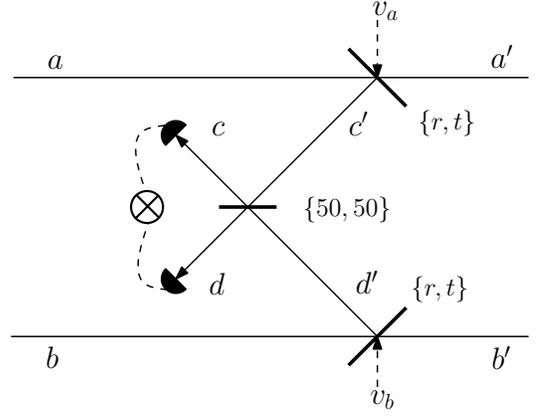} %
\caption{A single ``unit'': Fig.~\ref{Fig:whichway} plus an extra 50-50 beamsplitter which erases the which-way information and the relative phase is preserved at the output. The dashed lines connecting detectors $c$ and $d$ represents the fact that coincidence detection is only needed under certain circumstances which will be specified in later sections. On the other hand, the 50-50 beamsplitter is always necessary. } %
\label{Fig:qeraser} %
\end{figure}

\begin{table}[th]
\begin{center}
\begin{tabular}{ | c | c | c |}
    \hline
          detector state       &       transmitted state        & probability \\ \hline
     $\vert 0,0 \rangle_{d,c}$ &   $\frac{t^2}{\sqrt{2}} \left( \vert 2,0 \rangle_{a',b'} + e^{i 2 \phi} \vert 0,2 \rangle_{a',b'} \right)$   & $  t^4 $  \\ \hline
     $\vert 0,1 \rangle_{d,c}$ &   $-\frac{tr}{\sqrt{2}} \left( \vert 1,0 \rangle_{a',b'} -i e^{i 2 \phi}\vert 0,1 \rangle_{a',b'} \right)$    & $ t^2 r^2 $ \\ \hline
     $\vert 0,2 \rangle_{d,c}$ &   $\frac{r^2}{2\sqrt{2}}(1-e^{i 2 \phi})\vert 0,0 \rangle_{a',b'}$    & $ \frac{1}{2} r^4 \sin^2\phi $  \\ \hline
     $\vert 1,0 \rangle_{d,c}$ &   $\frac{itr}{\sqrt{2}} \left( \vert 1,0 \rangle_{a',b'} +i e^{i 2 \phi}\vert 0,1 \rangle_{a',b'} \right)$    & $ t^2 r^2 $ \\ \hline
     $\vert 2,0 \rangle_{d,c}$ &   $-\frac{r^2}{2\sqrt{2}}(1-e^{i 2 \phi})\vert 0,0 \rangle_{a',b'}$    & $ \frac{1}{2} r^4 \sin^2\phi $ \\ \hline
     $\vert 1,1 \rangle_{d,c}$ &   $-\frac{i r^2}{2}(1+e^{i 2 \phi})\vert 0,0 \rangle_{a',b'}$    & $  r^4 \cos^2\phi  $ \\ \hline
\end{tabular}
\caption{Probability of detecting all possible states at $c$ and $d$, for a $\vert 2::0 \rangle^{2\phi}$ input and erased which-way information, as shown in Fig.~\ref{Fig:qeraser}.}
\end{center}
\label{Tab:qeraser}
\end{table}

If an extra 50-50 beamsplitter is introduced as in Fig.~\ref{Fig:qeraser}, the information of which path (mode $c'$ or $d'$) the detected photons come from would be hidden from the environment. Therefore photons are subtracted coherently from the original state (modes $a$ and $b$) and the relative phase is preserved in the transmitted state (modes $a'$ and $b'$). The detection probabilities change accordingly and are shown in Table~II. Note that even though states such as $1/\sqrt{2} \left( \vert 1,0 \rangle_{a',b'} -i e^{i 2 \phi}\vert 0,1 \rangle_{a',b'} \right)$ and $1/\sqrt{2} \left( \vert 1,0 \rangle_{a',b'} +i e^{i 2 \phi}\vert 0,1 \rangle_{a',b'} \right)$ are single photon states and preserve the two-fold phase-resolving power, they are in general not valid output states because of the extra $i$ in the relative phase. However, we can apply certain phase shifts to make such states usable, which is discussed in later sections. Moreover, quantum interference is observed when all photons are reflected into the 50-50 beamsplitter and the corresponding probabilities vary as a function of the phase shift $\phi$, as would be expected from a lossless Mach-Zehnder interferometer. 

In conclusion, a ``quantum eraser'' is necessary to generate a super-resolving single-photon path-entangled state and we refer to the set-up shown in Fig.~\ref{Fig:qeraser} as a ``unit'', since it is used repeatedly in following calculation. In addition, with a general $\vert N::0 \rangle^{M\phi}$ as the input, the probability of detecting a $\vert m,n \rangle_{d,c}$ in a unit is shown to be
\begin{align}
\nonumber
&~P_{N}^{M\phi}(m,n,\rho) \\
\nonumber
=&~\vert_{d,c}\langle m,n \vert \psi_{\textrm{total}} \rangle \vert^2 \\
\nonumber
=&~\binom{N}{m+n}(1-\rho)^{N-(m+n)}\rho^{m+n} \binom{m+n}{m}\left(\frac{1}{2}\right)^{m+n}\\
&~\times \left\lbrace 1+\delta_{N,m+n} \cos\left(M\phi+\frac{(m-n)\pi}{2}\right) \right\rbrace.
\label{Eq:p(N,m,n)}
\end{align}
Here $\rho=r^2$, $1-\rho=t^2$ and $\vert \psi_{\textrm{total}}\rangle$ is the full output state in modes $a'$, $b'$, $d$ and $c$. In $P_{m,n}^N$, the expression $\binom{N}{m+n}(1-\rho)^{N-(m+n)}\rho^{m+n}$ comes from reflecting $m+n$ photons out of $N$ photons from the $\{ r, t \}$ beamsplitters while $\binom{m+n}{m}\left(\frac{1}{2}\right)^{m+n}$ is from the 50-50 beamsplitter. And the $\delta_{N,m+n}\times\cos$ term reflects the quantum interference when none of the $N$ photons is transmitted. A detailed derivation of Eq.~(\ref{Eq:p(N,m,n)}) is shown in Appendix~\ref{App:A}. 

%----------------------------------------------------------------------------------------

\section{Generation of $\mathbf{\vert 1::0 \rangle^{N \phi}}$ without number-resolving detectors} 

With ordinary photodetectors we are only able to tell if any photon arrives at the detector or not, but not how many of them. Consequently, it is unknown how many photons are reflected in a single unit. Therefore we introduce a stacking of units, in each of which the coincidence detection is implemented, to enhance the probability of generating the correct state.

To see how the coincidence detection is beneficial, we consider the example where the input of a unit is $\vert 3::0 \rangle^{3\phi}$. We take into account two probabilities. With the first probability, it is easy to see that there are ten possible detected states and only when a $\vert 1,1 \rangle_{d,c}$ is detected we generate the $\vert 1::0 \rangle_{a',b'}^{3\phi}$. This is a direct result from the Hong-Ou-Mandel (HOM) effect \cite{Hong1987,Lee2002}. With the second probability, if we do coincidence detection and take the output states only when both $c$ and $d$ click, the chance of $\vert 1,1 \rangle_{d,c}$ being detected is much higher, since there are only two other coincidence states: $\vert 2,1 \rangle_{d,c}$ and $\vert 1,2 \rangle_{d,c}$. We call the former the ``probability of detecting some state'' and the latter the ``conditional probability of detecting some state''. Both kinds of probabilities are discussed in following calculations.

\subsection{Odd photon number input state}

\begin{figure}[t] %
\centering %
\includegraphics[width=\linewidth]{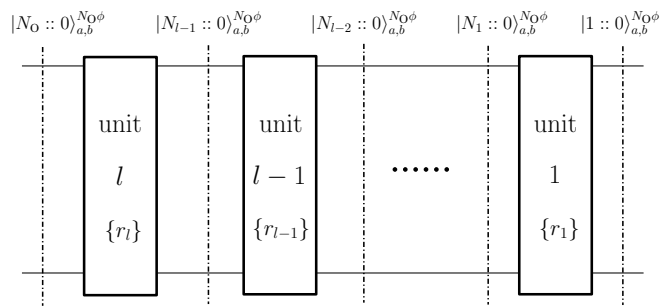} %
\caption{A stacking of units to reduce a $\vert N_{\textrm{o}} :: 0 \rangle^{N_{\textrm{o}} \phi}$, with $N_{\textrm{o}}=2 l+1$ being an odd number. Each rectangle represents the unit described in Fig.~\ref{Fig:qeraser}. Notice we count from the output for the ease of calculation. Unit $i$ with a reflectance of $r_i$ is the $i$th last unit before the output. It has a N00N state $\vert N_{i} ::0 \rangle^{N_{\textrm{o}} \phi}$ as the input state. Note that though the number of photon is decreasing from left to right, the relative phase is always $N_{\textrm{o}} \phi$, which is exactly what we need to generate a super-resolving single-photon path-entangled state.} 
\label{Fig:oddunits} %
\end{figure}

For a general input state $\vert N_{\textrm{o}} ::0 \rangle^{N_{\textrm{o}} \phi}$ with $N_{\textrm{o}}\equiv N_l=2 l +1$, $l=1,2,3...$, we need $l$ units to generate the desired output $\vert 1 ::0 \rangle^{N_{\textrm{o}} \phi}$, as shown in Fig.~\ref{Fig:oddunits}. The procedure is as follows: \textit{we propagate the input state from $l$th unit at the leftmost to the first unit right before the output, and for each unit $i$ we only pass on those cases where a coincidence between $d$ and $c$ is detected $($see Fig.~\ref{Fig:qeraser}$)$}. This makes sure that in each unit, at least two photons are subtracted from the input state while the multi-fold relative phase $e^{i N_{\textrm{o}}\phi}$ is intact. This is exactly what is needed to generate a super-resolving single-photon path-entangled state from a general maximally path-entangled state.

On the other hand, however, such a procedure has a low efficiency. The reason is that any state that loses more than three photons in any unit will NOT be propagated to the output because it has to lose one photon or none at all (i.e., no coincidence) in some other units to compensate for excess photon loss. Therefore there are only two possible output states: ($a$) $\vert 1::0 \rangle$, where every unit subtracts two photons, i.e., a $\vert 1,1 \rangle_{d,c}$ is detected in each unit; and ($b$) $\vert 0::0 \rangle$, where one of the units subtracts three photons, i.e., a $\vert 1,2 \rangle_{d,c}$ or a $\vert 2,1 \rangle_{d,c}$ is detected, and all others subtract two. 

For scenario ($a$), the probability can be calculated as
\begin{align}
\nonumber
P_{\text{all}~\vert1,1\rangle}^{N_{\textrm{o}}}=&~\displaystyle\prod_{k=1}^{l}P_{2k+1}^{N_{\textrm{o}}\phi}(1,1,\rho_k)\\
=&~\displaystyle\prod_{k=1}^{l}\frac{1}{2}k(2k+1)(1-\rho_k)^{2k-1}\rho_k^2.
\label{Eq:all11odd}
\end{align}
For scenario ($b$), we first show that the probability of detecting a $\vert 1,2 \rangle_{d,c}$ in the $i$th unit while a $\vert 1,1 \rangle_{d,c}$ is detected in all others is (a detailed derivation is presented in Appendix~\ref{App:B}.)
%\begin{widetext}
%\begin{align}
%\nonumber
%& P_{\text{$i$th}~
%\vert1,2\rangle}^{N_{\textrm{o}}} \\
%\nonumber
%=&~\left(\displaystyle\prod_{k=i+1}^{l}P_{2k+1}^{N_{\textrm{o}}\phi}(1,1,\rho_k)\right) P_{2i+1}^{N_{\textrm{o}}\phi}(1,2,\rho_i)\left(\displaystyle\prod_{k=1}^{i-1}P_{2k}^{N_{\textrm{o}}\phi}(1,1,\rho_k)\right)\\
%\nonumber
%=&~\left(\displaystyle\prod_{k=1}^{l}P_{2k+1}^{N_{\textrm{o}}\phi}(1,1,\rho_k)\right) \frac{P_{2i+1}^{N_{\textrm{o}}\phi}(1,2,\rho_i)}{P_{2i+1}^{N_{\textrm{o}}\phi}(1,1,\rho_i)} \left(\displaystyle\prod_{k=1}^{i-1}\frac{P_{2k}^{N_{\textrm{o}}\phi}(1,1,\rho_k)}{P_{2k+1}^{N_{\textrm{o}}\phi}(1,1,\rho_k)}\right)\\
%\nonumber
%=&~P_{\text{all}~\vert1,1\rangle}^{N_{\textrm{o}}} \frac{P_{2i+1}^{N_{\textrm{o}}\phi}(1,2,\rho_i)}{P_{2i+1}^{N_{\textrm{o}}\phi}(1,1,\rho_i)}\left(\displaystyle\prod_{k=1}^{i-1}\frac{P_{2k}^{N_{\textrm{o}}\phi}(1,1,\rho_k)}{P_{2k+1}^{N_{\textrm{o}}\phi}(1,1,\rho_k)}\right) \\
%=&~\begin{cases}
%P_{\text{all}~\vert1,1\rangle}^{N_{\textrm{o}}}\frac{\rho_1}{4(1-\rho_1)}(1+\sin(N_{\textrm{o}}\phi)), & i=1; \\
%P_{\text{all}~\vert1,1\rangle}^{N_{\textrm{o}}}\frac{\rho_i}{4(1-\rho_i)} (1+\cos(N_{\textrm{o}}\phi)) \times \displaystyle\prod_{k=1}^{i-1}\frac{1}{1-\rho_k} , & \text{else}.
%\end{cases}
%\label{Eq:one12}
%\end{align}
%\end{widetext}
\begin{align}
P_{\text{$i$th}~
\vert1,2\rangle}^{N_{\textrm{o}}}
=&~\begin{cases}
P_{\text{all}~\vert1,1\rangle}^{N_{\textrm{o}}} \frac{\rho_1 (1+\sin(N_{\textrm{o}}\phi))}{4(1-\rho_1)}, & i=1; \\
P_{\text{all}~\vert1,1\rangle}^{N_{\textrm{o}}} \frac{\rho_i (1+\cos(N_{\textrm{o}}\phi))}{4(1-\rho_i)}  \displaystyle\prod_{k=1}^{i-1}\frac{1}{1-\rho_k} , & \text{else}.
\end{cases}
\label{Eq:one12}
\end{align}
And the sum of all possible units for which $\vert 1,2 \rangle_{d,c}$ is detected becomes $P_{\text{one}~\vert1,2\rangle}^{N_{\textrm{o}}} = \sum_{i=1}^{l} P_{\text{$i$th}~\vert 1,2\rangle}^{N_{\textrm{o}}}$. Similarly, we have the probability of detecting a $\vert 2,1 \rangle_{d,c}$ in the $i$th unit while a $\vert 1,1 \rangle_{d,c}$ is detected in all others as
\begin{align}
P_{\text{$i$th}~\vert2,1\rangle}^{N_{\textrm{o}}} 
=~\begin{cases}
P_{\text{all}~\vert1,1\rangle}^{N_{\textrm{o}}} \frac{\rho_1 (1-\sin(N_{\textrm{o}}\phi))}{4(1-\rho_1)}, & i=1; \\
P_{\text{all}~\vert1,1\rangle}^{N_{\textrm{o}}} \frac{\rho_i (1+\cos(N_{\textrm{o}}\phi))}{4(1-\rho_i)} \displaystyle\prod_{k=1}^{i-1}\frac{1}{1-\rho_k} , & \text{else},
\end{cases}
\end{align}
and $P_{\text{one}~\vert2,1\rangle}^{N_{\textrm{o}}} = \sum_{i=1}^{l} P_{\text{$i$th}~\vert 2,1\rangle}^{N_{\textrm{o}}}$. Now the conditional probability of transmitting a $\vert 1::0 \rangle^{N_{\textrm{o}}\phi}$ after unit $1$, given the photon number of the input state is odd and coincidence detection is employed in every unit, becomes
\begin{align}
\nonumber
&~P_{\text{cond.}}^{N_{\textrm{o}}}\\
\nonumber
=&~\frac{P_{\text{all}~
\vert1,1\rangle}^{N_{\textrm{o}}}}{P_{\text{all}~
\vert1,1\rangle}^{N_{\textrm{o}}}+P_{\text{one}~
\vert1,2\rangle}^{N_{\textrm{o}}}+P_{\text{one}~
\vert2,1\rangle}^{N_{\textrm{o}}}} \\
=&~\frac{1}{\displaystyle\sum_{i=2}^{l} \frac{(1+\cos(N_{\textrm{o}}\phi))\rho_i}{2(1-\rho_i)} \displaystyle\prod_{k=1}^{i-1}\frac{1}{1-\rho_k} + \frac{2-\rho_1}{2(1-\rho_1)}}.
\label{Eq:oddcond}
\end{align}

The probability of generating $\vert 1::0 \rangle^{N_{\textrm{o}}}$ can be maximized by choosing a optimal reflectance $\rho_k$ for the $k$th unit. Physically, it is easy to show from Eq.~(\ref{Eq:all11odd}) that this probability maximizes when $\rho_k = 2/(2k+1)$ (i.e., the most probable number of photon being reflected in each unit is two) and
\begin{align}
\nonumber
P_{\text{all}~
\vert1,1\rangle}^{N_{\textrm{o}}}\vert_{\textrm{max}}=&~\prod_{k=1}^{l}\frac{2k}{2k-1}\left( \frac{2k-1}{2k+1} \right)^{2k} \\
=&~\frac{N_{\textrm{o}}!}{{N_{\textrm{o}}}^{N_{\textrm{o}}}}
\label{Eq:oddall11} \\
\approx&~\sqrt{2 \pi N_{\textrm{o}}} e^{-N_{\textrm{o}}},
\label{Eq:oddall11max}
\end{align} 
where in the last line Sterling's formula is used in large $N_{\textrm{o}}$ limit \cite{footnote1}. This probability decreases exponentially with increasing $N_{\textrm{o}}$ as expected. 

On the other hand, there is no optimal $\rho_k\in [0,1]$ that leads to a maximal conditional probability; $P_{\text{cond.}}^{N_{\textrm{o}}}$ in Eq.~(\ref{Eq:oddcond}) is independent of $P_{\text{all}~
\vert1,1\rangle}^{N_{\textrm{o}}}$ and approaches unity when all $\rho_k$ are close to zero. However, we can set a critical conditional probability $P_{\text{cond.}}^{N_{\textrm{o}}}\vert_{\textrm{c}}$ and compute the corresponding $\rho_k$. Assuming $N_\textrm{o}=7$, $\phi=\pi/14$, all $\rho_k$ are equal to $\rho_{\textrm{c}}$ and we want $50\%$ chance of transmitting a $\vert 1::0 \rangle^{7\phi}$ state whenever a coincidence is detected in unit $1$, i.e., $P_{\text{cond.}}^{N_{\textrm{o}}}\vert_{\textrm{c}}=0.5$, then $\rho_{\textrm{c}}$ can be calculated to be around $0.31$; when $P_{\text{cond.}}^{N_{\textrm{o}}}\vert_{\textrm{c}}=0.9$, $\rho_{\textrm{c}}\approx0.06$.  

However, this doesn't mean that we can obtain more outputs by decreasing the reflectance of each unit. In the previous example, the probability of generating $\vert 1::0 \rangle$ is $P_{\text{all}~\vert1,1\rangle}^{N_{\textrm{o}}}=0.2\%$ when $\rho=0.31$ in each unit and almost zero when $\rho=0.06$. Therefore, a higher conditional probability (i.e., higher fidelity) comes with a lower probability (i.e., efficiency), and vice versa. A similar reciprocal relation is described in Ref.~\cite{Kok2002}.

%-------------------------------------------------------------------------------------------------------------
\subsection{Even photon number input state}

For a general input state $\vert N_{\textrm{e}}::0 \rangle^{N_{\textrm{e}}\phi}$ with $N_{\textrm{e}}=2 l $, $l=1,2,3...$, we need $l$ units to generate the desired output $\vert 1::0 \rangle^{N_\textrm{e}\phi}$, as shown in Fig.~\ref{Fig:oddunits}. The procedure is as follows: \textit{we propagate the input state from $l$th unit at the leftmost to the second unit right before the first unit, and for each unit $i\in [2,l]$ we only pass on those cases where a coincidence between $d$ and $c$ $($see Fig.~\ref{Fig:qeraser}$)$ is detected; in the first unit however, we do a photo-detection on both $d$ and $c$ but no coincidence counting. Whenever $d$ $(c)$ clicks we apply a $-\pi/2$ $(\pi/2)$ phase shift on mode $b'$.} This makes sure that in each unit from $l$ to $2$, at least two photons are subtracted from the input state while the multi-fold relative phase $e^{i N_{\textrm{o}}\phi}$ is intact. 

Following a similar reasoning in the odd-number case, we can see that there are only two possible input states for unit $1$: ($a$) $\vert 2::0 \rangle$, where every unit from second to the $l$th subtracts two photons, i.e., a $\vert 1,1 \rangle_{d,c}$ is detected in each unit; and ($b$) $\vert 1::0 \rangle$, where one of the other units subtracts three photons, i.e., a $\vert 1,2 \rangle_{d,c}$ or a $\vert 2,1 \rangle_{d,c}$ is detected, and all others subtract two. Moreover, from Table~II, it is easy to see that a single click in $d$ corresponds to a $1/\sqrt{2}(\vert 1,0 \rangle_{a',b'}+ie^{i N_{\textrm{e}}\phi} \vert 0,1 \rangle_{a',b'})$ being transmitted, and a $-\pi/2$ phase shift on $b'$ turns it into a $\vert 1::0 \rangle^{N_{\textrm{e}}\phi}$. Similar logic applies to a single click in $c$. This justifies our procedure as a valid way of generating a super-resolving single-photon path-entangled state.

For scenario ($a$), the probability of having $\vert 2::0 \rangle^{N_{\textrm{e}}\phi}$ as the input of unit $1$ is 
\begin{align}
\nonumber
P_{\text{all}~\vert1,1\rangle}^{N_{\textrm{e}}}=&~\displaystyle\prod_{k=2}^{l}P_{2k}^{N_{\textrm{e}}\phi}(1,1,\rho_k)\\
=&~\displaystyle\prod_{k=2}^{l}\frac{1}{2}k(2k-1)(1-\rho_k)^{2(k-1)}\rho_k^2.
\label{Eq:all11even}
\end{align}
And the probability of detecting only one photon in unit 1 with this input state can be easily read off from Table~II as $2t_1^2r_1^2=2(1-\rho_1)\rho_1$. Therefore the probability of transmitting the correct state at the output is 
\begin{align}
2P_{\text{all}~\vert1,1\rangle}^{N_{\textrm{e}}}(1-\rho_1)\rho_1.
\label{Eq:evenp}
\end{align}

For scenario ($b$), we first show that the probability that a $\vert 1,2 \rangle_{d,c}$ is detected in the $i$th unit while a $\vert 1,1 \rangle_{d,c}$ is detected in all others is (a detailed derivation is presented in Appendix~\ref{App:B}.)
\begin{align}
P_{\text{$i$th}~
\vert1,2\rangle}^{N_{\textrm{e}}}
=~P_{\text{all}~\vert1,1\rangle}^{N_{\textrm{e}}}\frac{\rho_i}{2(1-\rho_i)}\displaystyle\prod_{k=2}^{i-1}\frac{1}{1-\rho_k}.
\label{Eq:evenith}
\end{align}

And the sum of all possible units for which a $\vert 1,2 \rangle_{d,c}$ is detected becomes $P_{\text{one}~\vert1,2\rangle}^{N_{\textrm{e}}} = \sum_{i=2}^{l} P_{\text{$i$th}~\vert 1,2\rangle}^{N_{\textrm{e}}}$. Because of symmetry, we have $P_{\text{$i$th}~\vert2,1\rangle}^{N_{\textrm{e}}} = P_{\text{$i$th}~
\vert1,2\rangle}^{N_{\textrm{e}}}$ and $P_{\text{one}~\vert 2,1 \rangle}^{N_{\textrm{e}}}=P_{\text{one}~\vert 1,2 \rangle}^{N_{\textrm{e}}}$. 

Now we are in the position of calculating the conditional probability of transmitting a $\vert 1::0 \rangle^{N_{\textrm{e}}\phi}$ in the case of even photon number input state, given a coincidence is detected in unit $2$ to unit $l$ and either $d$ or $c$ (but not both) in unit $1$ clicks:
\begin{widetext}
\begin{align}
\nonumber
P_{\text{cond.}}^{N_{\textrm{e}}} =&~\frac{2 P_{\text{all}~\vert1,1\rangle}^{N_{\textrm{e}}}(1-\rho_1)\rho_1}
{P_{\text{all}~\vert1,1\rangle}^{N_{\textrm{e}}}\left\lbrace 2-\rho_1\left(1+\cos^2 \left(\frac{N_{\textrm{e}}}{2}\phi \right)\right)\right\rbrace\rho_1+\left(P_{\text{one}~
\vert1,2\rangle}^{N_{\textrm{e}}}+P_{\text{one}~
\vert2,1\rangle}^{N_{\textrm{e}}}\right)\rho_1} \\
 =&~ \frac{2(1-\rho_1)}{2-\rho_1\left(1+\cos^2 \left(\frac{N_{\textrm{e}}}{2}\phi \right)\right) + \displaystyle \sum_{i=2}^{l} \frac{\rho_i}{1-\rho_i} \displaystyle\prod_{k=2}^{i-1}\frac{1}{1-\rho_k}}.
\label{Eq:evencond}
\end{align}
\end{widetext}
The $\left\lbrace 2-\rho_1\left(1+\cos^2 \left(N_{\textrm{e}}\phi/2 \right)\right)\right\rbrace\rho_1$ in the denominator represents the probability of getting one click in unit $1$ given its input is a $\vert 2::0 \rangle^{N_{\textrm{e}}\phi}$, while $\rho_1$ represents that of a input state of $\vert 1::0 \rangle^{N_{\textrm{e}}\phi}$.

From Eqs.~(\ref{Eq:all11even}) and (\ref{Eq:evenp}), the probability of generating $\vert 1::0 \rangle^{N_{\textrm{e}}\phi}$ can be maximized by choosing an optimal reflectance $\rho_k$ for the $k$th unit. Physically, it is easy to show that $P_{\text{all}~\vert1,1\rangle}^{N_{\textrm{e}}}$ maximizes when $\rho_k = 1/k$ and
\begin{align}
\nonumber
P_{\text{all}~
\vert1,1\rangle}^{N_{\textrm{e}}}\vert_{\textrm{max}}=&~\prod_{k=2}^{l}\frac{2k-1}{2k}\left( \frac{k-1}{k} \right)^{2(k-1)} \\
=&~2 \frac{N_{\textrm{e}}!}{{N_{\textrm{e}}}^{N_{\textrm{e}}}} 
\label{Eq:evenall11} \\
\approx&~2 \sqrt{2 \pi N_{\textrm{e}}} e^{-N_{\textrm{e}}},
\label{Eq:evenall11max}
\end{align} 
where in the last line Sterling's formula is used in large $N_{\textrm{e}}$ limit. And it is easy to show the expression $2(1-\rho_1)\rho_1$ in Eq.~(\ref{Eq:evenp}) has a maximal value of $1/2$ when $\rho_1=1/2$. Therefore the maximum probability of transmitting the correct state is $N_{\textrm{e}}!/{N_{\textrm{e}}}^{N_{\textrm{e}}}$, which agrees with Eq.~(\ref{Eq:oddall11}) and can be reached when the reflectance of the units are such that
\begin{align}
\rho_k=
\begin{cases}
1/k, & k\in [2,N_{\textrm{e}}/2];\\
1/2, & k=1.
\end{cases}
\end{align}

On the other hand, there is no optimal $\rho_k\in [0,1]$ that leads to a maximal conditional probability in Eq.~(\ref{Eq:evencond}). Just as the case with odd photon number input state, we may set a critical probability and ask the corresponding $\rho_k$, and the reciprocal relation between fidelity and efficiency stands as well.

%------------------------------------------------------------------------------------------------------------
\section{Generation of $\mathbf{\vert 1::0 \rangle^{N \phi}}$ with number-resolving detectors}

With number-resolving detectors implemented at $d$ and $c$ in Fig.~\ref{Fig:qeraser}, we are able to tell how many photons are reflected in a unit \cite{Kim1999, Takeuchi1999}. Therefore it is much easier to generate a $\vert 1::0 \rangle^{N\phi}$ from a general $\vert N::0 \rangle^{N\phi}$---only a single unit is needed. The protocol is as follows: \textit{assuming $m$ and $n$ photons are detected at $d$ and $c$ in a single unit with reflectance $\rho$, we $(a)$ propagate the state to the output only when $m+n=N-1$, and $(b)$ given $(a)$ is true, we apply a $(n-m)\pi/2$ phase shift on mode $b'$.} This protocol can be easily derived.

A general input N00N state in Fig.~\ref{Fig:qeraser} is 
\begin{align}
\vert \psi_{\textrm{in}} \rangle 
=\frac{1}{\sqrt{2}}  \left( \vert N,0;0,0 \rangle_{a,v_a;b,v_b} + e^{i N\phi} \vert 0,0;N,0 \rangle_{a,v_a;b,v_b} \right),
\end{align}
where $N$ can be either odd or even. The corresponding output state at $a'$, $b'$, $c'$ and $d'$ is
\begin{widetext}
\begin{align}
\vert \psi_{\textrm{out}} \rangle 
= \displaystyle\sum\limits_{k=0}^N \binom{N}{k}^{1/2} t^k (i r)^{N-k} \times 
\frac{1}{\sqrt{2}} ( \vert k,0;N-k,0 \rangle_{a',b';c',d'} + e^{i N \phi}  \vert 0,k;0,N-k \rangle_{a',b';c',d'} ).
\label{Eq:output}
\end{align}
Given we are only interested in transmitting one photon to the output, states in $\vert \psi_{\textrm{out}} \rangle$ with $k\neq 1$ can be ignored and we are left with
\begin{align}
\vert \psi_{\textrm{target}} \rangle 
=& \frac{\sqrt{N}}{\sqrt{2}} t (ir)^{N-1}  \left( \vert 1,0 \rangle_{a',b'} \otimes \vert N-1,0 \rangle_{c',d'} + e^{i N \phi}\vert 0,1 \rangle_{a',b'} \otimes \vert 0,N-1 \rangle_{c',d'} \right). 
\label{Eq:target}
\end{align}
Since $\vert k,0 \rangle_{c',d'} \rightarrow \left(\frac{1}{\sqrt{2}} \right)^{k} \sum\limits_{m=0}^{k} \binom{k}{m}^{\frac{1}{2}} i^{n} \vert m,n \rangle_{d,c}$ and $\vert 0,k \rangle_{c',d'} \rightarrow  \left(\frac{1}{\sqrt{2}} \right)^{k} \sum\limits_{m=0}^{k} \binom{k}{m}^{\frac{1}{2}} i^{m} \vert m,n \rangle_{d,c}$, with $k=N-1$ Eq.~(\ref{Eq:target}) becomes
\begin{align}
\vert \psi_{\textrm{target}} \rangle 
=& \sqrt{N} t \left(\frac{ir}{\sqrt{2}}\right)^{N-1} \sum\limits_{m=0}^{N-1} \binom{N-1}{m}^{\frac{1}{2}} i^{n} \left( \frac{\vert 1,0 \rangle_{a',b'}  + i^{m-n} e^{i N \phi}\vert 0,1 \rangle_{a',b'}}{\sqrt{2}} \right) \otimes \vert m,n \rangle_{d,c}.
\end{align}
\end{widetext}
The state in the last parentheses is a $\vert 1::0\rangle^{N\phi}$ with an extra phase of $(m-n)\pi/2$, which can be corrected by applying an $(n-m)\pi/2$ phase shift on mode $b'$. This allows us to keep all transmitted states with such detected photon numbers $m$ and $n$ in mode $d$ and $c$ that $m+n = N-1$. In contrary, there are always faulty transmitted states when only coincidence detection is available. 

\begin{figure}[h] %
\centering %
\includegraphics[width=1\linewidth]{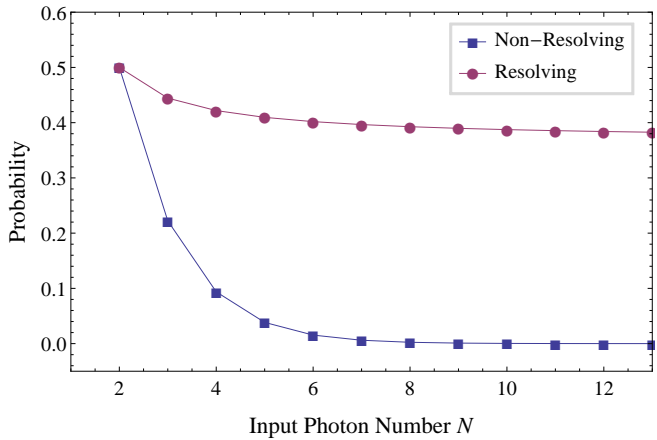} %
\caption{The maximized probabilities of transmitting a $\vert 1::0 \rangle^{N\phi}$ state following different protocols (see Eqs.~(\ref{Eq:oddall11}), (\ref{Eq:evenall11}) and (\ref{Eq:resolving})) are plotted as a function of the input photon number. The squares represent the probabilities for a general input state as in Eq.~(\ref{Eq:N00M}) without photon number resolving detections while the disks represents the probabilities with photon number-resolving detections. It is clear that with increasing input photon number, number resolving detection maintains around 40$\%$ efficiency while coincidence detection drops to zero rapidly.} %
\label{Fig:comparison} %
\end{figure}

From Eqs.~(\ref{Eq:output}) and (\ref{Eq:target}), the probability of transmitting the correct state with photon number-resolving detectors is 
\begin{align}
P_{\textrm{resolving}}^N=\vert \langle \psi_{\textrm{target}} \vert \psi_{\textrm{out}} \rangle \vert^2=N(1-\rho)\rho^{N-1},
\end{align}
where $\rho=r^2$ as before. This probability maximizes at $\rho=r^2=(N-1)/N$, with
\begin{align}
P_{\textrm{resolving}}^N\vert_{\textrm{max}}=&\left( \frac{N-1}{N} \right)^{N-1} 
\label{Eq:resolving} \\
\approx& e^{-1},
\end{align}
where in the last line we take the limit of large $N$.

In Fig.~\ref{Fig:comparison} we plot the probabilities for protocols with and without number-resolving detectors. It is easy to see that with photon number-resolving detectors, a  super-resolving single-photon path-entangled state is much more likely to be generated from a general N00N state. Moreover, since we have complete control over the state being transmitted, the conditional probability with photon number-resolving detectors is always one. 
%-----------------------------------------------------------------------------------

\section{Conclusion and Discussion}
In this paper, we have proposed two protocols for generating a super-resolving single-photon path-entangled state $\vert 1::0 \rangle^{N\phi}$ from a general $N$ photon N00N state, either with or without photon number resolving detectors. On one hand, both protocols require an extra 50-50 beamsplitter on top of a general two-path interferometer to cloak the which-way information of the reflected photons to maintain the coherence of the transmitted states. On the other hand, the generation of $\vert 1::0 \rangle^{N\phi}$ without photon number resolving detectors has to be realized in $\lfloor{N/2}\rfloor$ steps, whereas that with photon number resolving detectors can be done in one step since we have full knowledge of the detected photon numbers. In addition, we have shown that both the probability (efficiency) and conditional probability (fidelity) of generation are higher when photon number resolving detectors are involved. In the case with non-resolving detectors we can achieve arbitrarily high fidelity, however at the cost of low efficiency. 

We conclude this paper with a discussion on the effect of the imperfection of detectors. Following the same argument in Ref.~\cite{Kok2002} we may ignore the deteriorated efficiency and dark counts because of the short operation-time windows, and consider only the imperfect detection efficiency of the detectors. First we consider the case without photon number resolving detectors and assume all detectors are the same. From Ref.~\cite{footnote1} it is easy to see that the numerator of Eq.~(\ref{Eq:oddall11}) now becomes $N_{\textrm{o}}!\eta^{N_{\textrm{o}}}$, where $\eta$ is the detector efficiency and $\eta^{N_{\textrm{o}}}$ is the probability that each of $N_{\textrm{o}}$ detectors clicks accurately. Thus this protocol performs exponentially poorly when detector efficiency is not unity. With photon number resolving detectors, since only two detectors (one unit) are involved, the effect of detector efficiency scales as $\eta^2$, which is independent of the number of input photon. 

\section*{Acknowledgements}
The research of W. F. is supported by the National Basic Research Program of China and the National Natural Science Foundation of China. K. J. and J. P. D. would like to acknowledge support from the Air Force Office of Scientific Research, the Army Research Office, and the National Science Foundation. M. L.-J. L. would like to acknowledge support from the National Science Foundation. The research of M. S. Z. is supported by NPRP grant (NPRP 5-102-1-026) from the Qatar National Research Fund (QNRF). We thank S.-Y. Zhu for interesting and useful discussions. 

%-----------------------------------------------------------------------------------

\appendix
\section{Derivation of Eq.~(\ref{Eq:p(N,m,n)})}
\label{App:A}
For a general input state as in Eq.~(\ref{Eq:N00M}), the output state in modes $a'$, $b'$, $d$ and $c$ is
\begin{widetext}
\begin{align}
\nonumber
\vert \psi_{\textrm{total}} \rangle =&~ \frac{1}{\sqrt{2}} \left( \displaystyle\sum\limits_{k=0}^N \sqrt{\binom{N}{k}}
t^{N-k} (i r)^k \vert N-k,0 \rangle_{a',b'} \otimes \left(\frac{1}{\sqrt{2}} \right)^{k} \displaystyle\sum\limits_{l=0}^k \sqrt{\binom{k}{l}} i^{k-l} \vert l,k-l \rangle_{d,c} \right) \\
\nonumber
&~+ e^{i M \phi} \frac{1}{\sqrt{2}} \left( \displaystyle\sum\limits_{k=0}^N \sqrt{\binom{N}{k}}
t^{N-k} (i r)^k \vert 0,N-k \rangle_{a',b'} \otimes \left(\frac{1}{\sqrt{2}} \right)^{k} \displaystyle\sum\limits_{l=0}^k \sqrt{\binom{k}{l}} i^{k-l} \vert k-l,l \rangle_{d,c} \right) \\
\nonumber
=&~ \displaystyle\sum\limits_{k=0}^N \sqrt{\binom{N}{k}} t^{N-k} \left(\frac{ir}{\sqrt{2}} \right)^{k} \\
&~\times \frac{1}{\sqrt{2}} \Bigg\lbrace  \vert N-k,0 \rangle_{a',b'} \otimes \displaystyle\sum\limits_{l=0}^k \sqrt{\binom{k}{l}} i^{k-l} \vert l,k-l \rangle_{d,c} 
+ e^{i M \phi} \vert 0,N-k \rangle_{a',b'} \otimes \displaystyle\sum\limits_{l=0}^k \sqrt{\binom{k}{l}} i^{l} \vert l,k-l \rangle_{d,c}    \Bigg\rbrace.
\end{align}

The probability of detecting an arbitrary state $\vert m,n \rangle_{d,c}$ is then
\begin{align}
\nonumber
P_{m,n}^{N}=&~\left\vert_{d,c}\langle m,n \vert \psi_{\textrm{total}} \rangle \right\vert^2 \\
\nonumber
=&~\Bigg\vert \sqrt{\binom{N}{m+n}} t^{N-(m+n)} \left(\frac{ir}{\sqrt{2}} \right)^{m+n} \\
\nonumber
&~\times \frac{1}{\sqrt{2}} \Bigg\lbrace  \vert N-(m+n),0 \rangle_{a',b'} \otimes \sqrt{\binom{m+n}{m}} i^{n}  + e^{i M \phi} \vert 0,N-(m+n) \rangle_{a',b'} \otimes \sqrt{\binom{m+n}{m}} i^{m}  \Bigg\rbrace  \Bigg\vert^2 \\
\nonumber
=&~\left(\frac{1}{2}\right)^{m+n}\binom{N}{m+n}\binom{m+n}{m}(1-\rho)^{N-(m+n)}\rho^{m+n} \\
\nonumber
&~\times \left\vert \frac{1}{\sqrt{2}} \Bigg\lbrace i^{n} \vert N-(m+n),0 \rangle_{a',b'}  + i^{m} e^{i M \phi} \vert 0,N-(m+n) \rangle_{a',b'}   \Bigg\rbrace \right\vert^2 \\
=&~\binom{N}{m+n}\tau^{N-(m+n)}\rho^{m+n} \binom{m+n}{m}\left(\frac{1}{2}\right)^{m+n} \left\lbrace 1+\delta_{N,m+n} \cos\left(M\phi+\frac{(m-n)\pi}{2}\right) \right\rbrace.
\end{align}
\end{widetext}

\section{Derivation of Eqs.~(\ref{Eq:one12}) and (\ref{Eq:evenith})}
\label{App:B}
For scenario ($b$) in the case of odd photon number input state, the probability of detecting a $\vert 1,2 \rangle_{d,c}$ in the $i$th unit while a $\vert 1,1 \rangle_{d,c}$ is detected in all others is
\begin{widetext}
\begin{align}
\nonumber
P_{\text{$i$th}~
\vert1,2\rangle}^{N_{\textrm{o}}} 
=&~\left(\displaystyle\prod_{k=i+1}^{l}P_{2k+1}^{N_{\textrm{o}}\phi}(1,1,\rho_k)\right) P_{2i+1}^{N_{\textrm{o}}\phi}(1,2,\rho_i)\left(\displaystyle\prod_{k=1}^{i-1}P_{2k}^{N_{\textrm{o}}\phi}(1,1,\rho_k)\right)\\
\nonumber
=&~\left(\displaystyle\prod_{k=1}^{l}P_{2k+1}^{N_{\textrm{o}}\phi}(1,1,\rho_k)\right) \frac{P_{2i+1}^{N_{\textrm{o}}\phi}(1,2,\rho_i)}{P_{2i+1}^{N_{\textrm{o}}\phi}(1,1,\rho_i)} \left(\displaystyle\prod_{k=1}^{i-1}\frac{P_{2k}^{N_{\textrm{o}}\phi}(1,1,\rho_k)}{P_{2k+1}^{N_{\textrm{o}}\phi}(1,1,\rho_k)}\right)\\
\nonumber
=&~P_{\text{all}~\vert1,1\rangle}^{N_{\textrm{o}}} \frac{P_{2i+1}^{N_{\textrm{o}}\phi}(1,2,\rho_i)}{P_{2i+1}^{N_{\textrm{o}}\phi}(1,1,\rho_i)}\left(\displaystyle\prod_{k=1}^{i-1}\frac{P_{2k}^{N_{\textrm{o}}\phi}(1,1,\rho_k)}{P_{2k+1}^{N_{\textrm{o}}\phi}(1,1,\rho_k)}\right) \\
=&~\begin{cases}
P_{\text{all}~\vert1,1\rangle}^{N_{\textrm{o}}}\frac{\rho_1}{4(1-\rho_1)}(1+\sin(N_{\textrm{o}}\phi)), & i=1; \\
P_{\text{all}~\vert1,1\rangle}^{N_{\textrm{o}}}\frac{\rho_i}{4(1-\rho_i)} (1+\cos(N_{\textrm{o}}\phi)) \times \displaystyle\prod_{k=1}^{i-1}\frac{1}{1-\rho_k} , & \text{else}.
\end{cases}
\end{align}

Similarly, for scenario ($b$) in the case of even photon number input state, the probability of detecting a $\vert 1,2 \rangle_{d,c}$ in the $i$th unit while a $\vert 1,1 \rangle_{d,c}$ is detected in all others is

\begin{align}
\nonumber
P_{\text{$i$th}~
\vert1,2\rangle}^{N_{\textrm{e}}} =&~\left(\displaystyle\prod_{k=i+1}^{l}P_{2k}^{N_{\textrm{e}}\phi}(1,1,\rho_k)\right)\times P_{2i}^{N_{\textrm{e}}\phi}(1,2,\rho_i)\times
\left(\displaystyle\prod_{k=2}^{i-1}P_{2k-1}^{N_{\textrm{e}}\phi}(1,1,\rho_k)\right)\\
\nonumber
=&~\left(\displaystyle\prod_{k=2}^{l}P_{2k}^{N_{\textrm{e}}\phi}(1,1,\rho_k)\right) \times \frac{P_{2i}^{N_{\textrm{e}}\phi}(1,2,\rho_i)}{P_{2i}^{N_{\textrm{e}}\phi}(1,1,\rho_i)}\times \left(\displaystyle\prod_{k=2}^{i-1}\frac{P_{2k-1}^{N_{\textrm{e}}\phi}(1,1,\rho_k)}{P_{2k}^{N_{\textrm{e}}\phi}(1,1,\rho_k)}\right)\\
\nonumber
=&~P_{\text{all}~\vert1,1\rangle}^{N_{\textrm{e}}} \times \frac{P_{2i}^{N_{\textrm{e}}\phi}(1,2,\rho_i)}{P_{2i}^{N_{\textrm{e}}\phi}(1,1,\rho_i)}\times \left(\displaystyle\prod_{k=2}^{i-1}\frac{P_{2k-1}^{N_{\textrm{e}}\phi}(1,1,\rho_k)}{P_{2k}^{N_{\textrm{e}}\phi}(1,1,\rho_k)}\right)\\
=&~P_{\text{all}~\vert1,1\rangle}^{N_{\textrm{e}}}\times \frac{\rho_i}{2(1-\rho_i)} \times \displaystyle\prod_{k=2}^{i-1}\frac{1}{1-\rho_k}.
\end{align}
\end{widetext}

\bibliography{reference}

\begin{thebibliography}{22}
\bibitem [{foo()}]{footnote1}%
  \BibitemOpen
  \href@noop {} {}\bibinfo {note} {In the general case, for each of the
  $N_{\textrm{o}}$ photons there are $2l+1$ possible ways of being reflected
  (there are two paths in each of the $l$ units and one accounts for the case
  where no reflection occurs), which explains the denominator in
  Eq.~(\ref{Eq:oddall11}). On the other hand, when only two photons are
  expected to be reflected in a unit with a input state $\vert N::0 \rangle$,
  the number of possible ways of such reflection is $2\binom{N}{2}$. Therefore
  the total number of possible reflections in this case is
  $\prod_{k=1}^{l}2\binom{N_{k}}{2}=\prod_{k=1}^{N_{\textrm{o}}}\binom{k}{1}=N_{\textrm{o}}!$,
  which is the numerator of Eq.~(\ref{Eq:oddall11}).}\BibitemShut {Stop}%
\end{thebibliography}
\end{document}